# The metallic resistance of a dilute two-dimensional hole gas in a GaAs quantum well: two-phase separation at finite temperature?

Xuan P. A. Gao[1,2], Allen P. Mills, Jr.[1,3], Arthur P. Ramirez[4], Loren N. Pfeiffer[1] and Kenneth W. West[1]
[1]*Bell Laboratories, Lucent Technologies, Murray Hill, New Jersey 07974*
[2]*Dept. of Applied Physics & Applied Math, Columbia University, New York City, NY 10027*
[3]*Dept. of Physics, University of California, Riverside, CA 92521*
[4]*Los Alamos National Laboratory, Los Alamos, NM 87545*

We have studied the magnetotransport properties of a high mobility two-dimensional hole gas (2DHG) system in a 10nm GaAs quantum well (QW) with densities in range of 0.7-1.6×10 $^{10}$ cm $^{-2}$ on the metallic side of the zero-field 'metal-insulator transition' (MIT). In a parallel field well above $B_c$ that suppresses the metallic conductivity, the 2DHG exhibits a conductivity $g(T) \approx 0.3(e^2/h)\ln T$ reminiscent of weak localization. The experiments are consistent with the coexistence of two phases in our system: a metallic phase and a weakly insulating Fermi liquid phase having a percolation threshold close to $B_c$.

Since the first report of a metal-insulator transition (MIT) at zero magnetic field in a 2D electron gas (2DEG) in a clean Si-MOSFET, the nature of the transition and the origin of the metallic-like behavior have been a subject of much debate [1]. The destruction of the anomalous metallic state and the large magneto-resistance induced by an in-plane magnetic field ($B_{\|}$) have suggested that the spins of the carriers are decisive in the underlying mechanism responsible for the 2D metallic conduction [1,2]. Recently the scaling of the magnetoconductivity of 2DEG in Si-MOSFET's for a wide range of densities and temperatures has been taken as evidence for a quantum phase transition in those systems [3]. In Ref. 3 it is further concluded that the MIT is related to a ferromagnetic instability of the 2DEG in Si-MOSFET's. A similar conclusion was reached independently because of the observation of the vanishing of the magnetic field required to fully polarize the 2DEG at the critical density of the zero field MIT [4]. These observations further stress the importance of spin interactions in the microscopic mechanism responsible for the metallic phenomenon in Si-MOSFET.

In a previous study, we found that the activation energy $E_a$, which characterizes the Arrhenius temperature dependent metallic resistance drop of the 2DHG in GaAs QW's, is insensitive to the in-plane magnetic field [5]. We found in Ref. 5 that an in-plane field less than the critical field only suppresses the magnitude of the resistance drop without affecting $E_a$. This fact hints that spin interactions may be irrelevant to the origin of the thermally activated resistance drop of the 2DHG in GaAs/AlGaAs QW's. One of the few viable explanations for the activated resistance is the mixed phase percolation-driven MIT scenario of He and Xie involving the condensation of liquid droplets from a low density phase at a finite temperature [6,7]. In this letter we show that, the phase separation model is consistent with our data, and the low density phase is behaving like a weakly localized ordinary Fermi liquid.

In this letter, we report our transport measurements over an extended density range on a dilute 2DHG system when a parallel magnetic field is applied along the [01$\underline{1}$] or [$\underline{2}$33] crystallographic directions. The critical field $B_c$, which fully suppresses the resistance drop, is found to be a linear function of $p$, the density of 2DHG and extrapolates to zero at $p_c$, the critical density of the zero field MIT similar to previously reported [8]. The values of $B_c$ with $B_{\|}$ applied along [01$\underline{1}$] or [$\underline{2}$33] are found here to be quantitatively correlated with the anisotropic $g$ factor in GaAs QW's over all the density range studied. It is observed that in the high temperature range before the resistance drops, the parallel field magnetoresistance is negligibly small. As the metallic resistance begins to appear below a temperature $T_0$, a parallel field less than $B_c$ only suppresses the size of resistance drop without affecting the characteristic energy scale, as previously reported [5]. When $B_{\|}$ is well above $B_c$, a logarithmic temperature dependent conductivity reminiscent of weak localization is observed over a wide range of temperature, density and resistivity (strength of disorder) when the conductivity $g$ is greater than $e^2/h$. These results agree with a two-phase coexistence picture in the anomalous 2D metallic state of our system: a metallic liquid phase coexists with a weakly localized 'normal' Fermi liquid gas phase. Under this picture, the thermally activated metallic resistance could be due to the formation of liquid droplets at finite temperature [6,7]. The density dependence of the anisotropic $B_c$ can also be readily understood in terms of a constant $g$ factor over all the density range studied down to $p_c$, which is in variance with the ferromagnetic instability observed for electrons in Si-MOSFET's [3,4].

Our transport measurements were performed on a high mobility low density 2DHG in a 10nm wide GaAs quantum well. The measurement driving signals were limited below 3fW per cm$^2$ sample area to avoid heating the 2DHG down to 10mK. The 2DHG has a density of $1.14\times10^{10}$ cm$^{-2}$ from doping, and a low temperature hole mobility of $3.4\times10^5$cm$^2$ V$^{-1}$ s$^{-1}$ without gating. This very same sample was previously used in [5]. The sample was grown on a (311)A GaAs wafer using $Al_xGa_{1-x}As$ barriers (typical x = 0.10). Delta-doping layers of Si dopants were symmetrically placed above and below the pure GaAs QW in order to minimize the asymmetry-induced spin nondegeneracy. The sample was prepared in the form of a Hall bar, of approximate dimensions (2.5×9) mm$^2$, with diffused In(5% Zn) contacts. The density of the 2DHG was tuned by a metallic back gate, which is about 100 microns beneath the well. The sample exhibits an apparent MIT at a critical density $p_c\sim0.6\times10^{10}$cm$^{-2}$, the lowest critical density in similar structures found to date by our knowledge. In this study we report our investigation on the metallic side of the MIT over the density range of $p$=0.7-1.6×10$^{10}$ cm$^{-2}$. This corresponds to a Wigner-Seitz radius $r_s$ = 12-18, with hole effective band mass taken as $0.18m_e$ [9]. The measurement current (~ 100pA, 4Hz square wave) was applied along the [$\underline{2}$33] direction in all our experiments. Independent measurements of the longitudinal resistance per square, $R_{xx}$, from contacts on both sides of the sample were made simultaneously as the temperature or applied magnetic field was varied. The samples were mounted in a top-loading dilution refrigerator. The temperature was read from a Ge resistance thermometer attached to the refrigerator mixing chamber. The Ge thermometer was calibrated down to 4mK by He-3 melting curve thermometry.

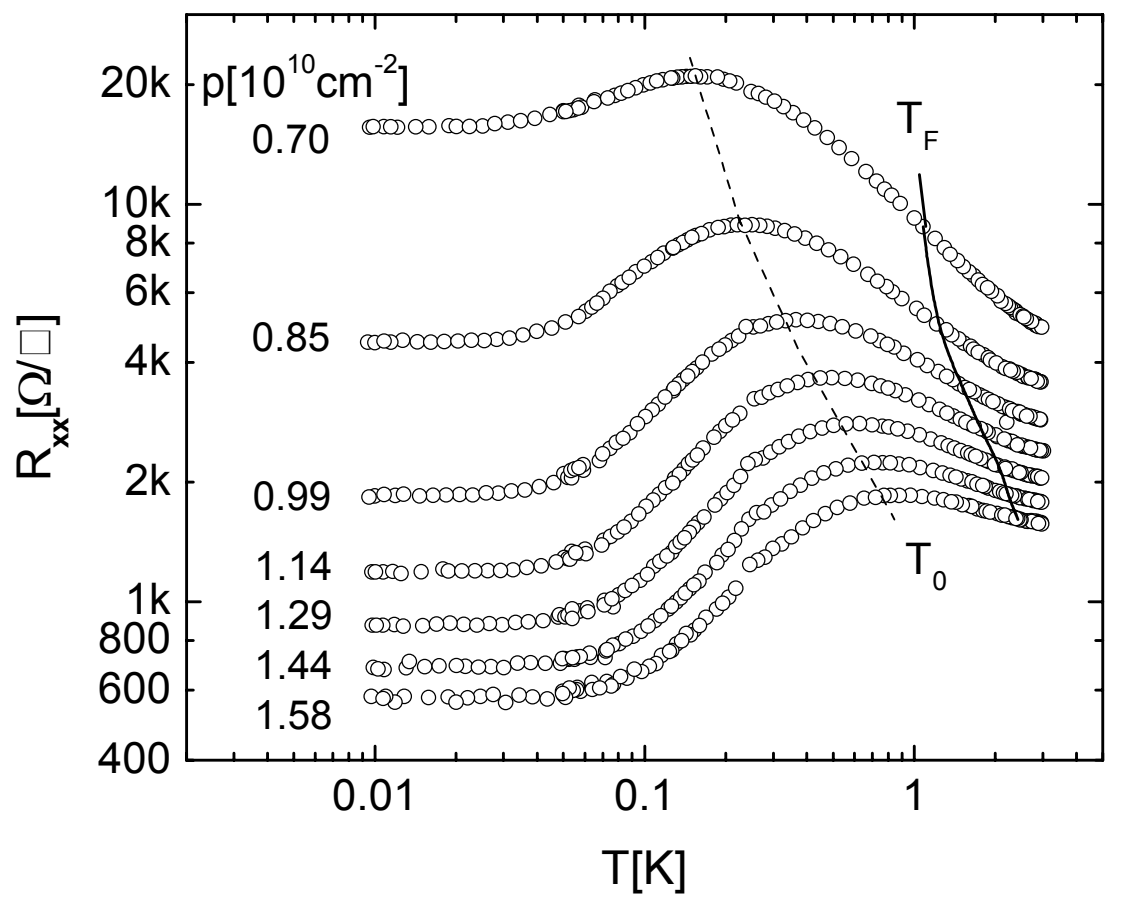

FIG. 1. Resistance per square, $R_{xx}$ vs. T for 2DHG with various densities in a 10nm wide GaAs quantum well at zero magnetic field. The black solid line marks corresponding $T_F$'s, the Fermi temperatures of the 2DHG's. The dashed black line connects the characteristic temperature $T_0$, below which the system exhibits a thermally activated metallic resistance.

The zero magnetic field temperature dependence of $R_{xx}$ for the sample with different 2DHG densities is shown in Fig. 1 from 10mK to 3K. All the $R_{xx}(T)$ traces exhibit a nonmonotonic peak, which is frequently observed in high mobility GaAs/AlGaAs heterostructures or quantum wells [10,11]. It can be seen that for $T\sim T_F$, $R_{xx}(T)$ is insulating like, i.e. d$R_{xx}(T)$/d$T$<0. This type of increasing resistivity upon lowering $T$ at temperatures around $T_F$ has been understood as the quantum-classical crossover when the 2D system becomes degenerate [12]. As $T<T_0$, a thermally activated metallic-like $R_{xx}(T)$ shows up. The origin of this thermally activated resistance has been the center of debate over almost a decade [1]. Similar to earlier reports [10,11], it is also obvious from Fig. 1 that $T_0$ roughly follows the $T_F$ (or density) of the 2D system, which suggests the metallic behavior for $T<T_0$ is associated with the quantum degeneracy of the 2DHG.

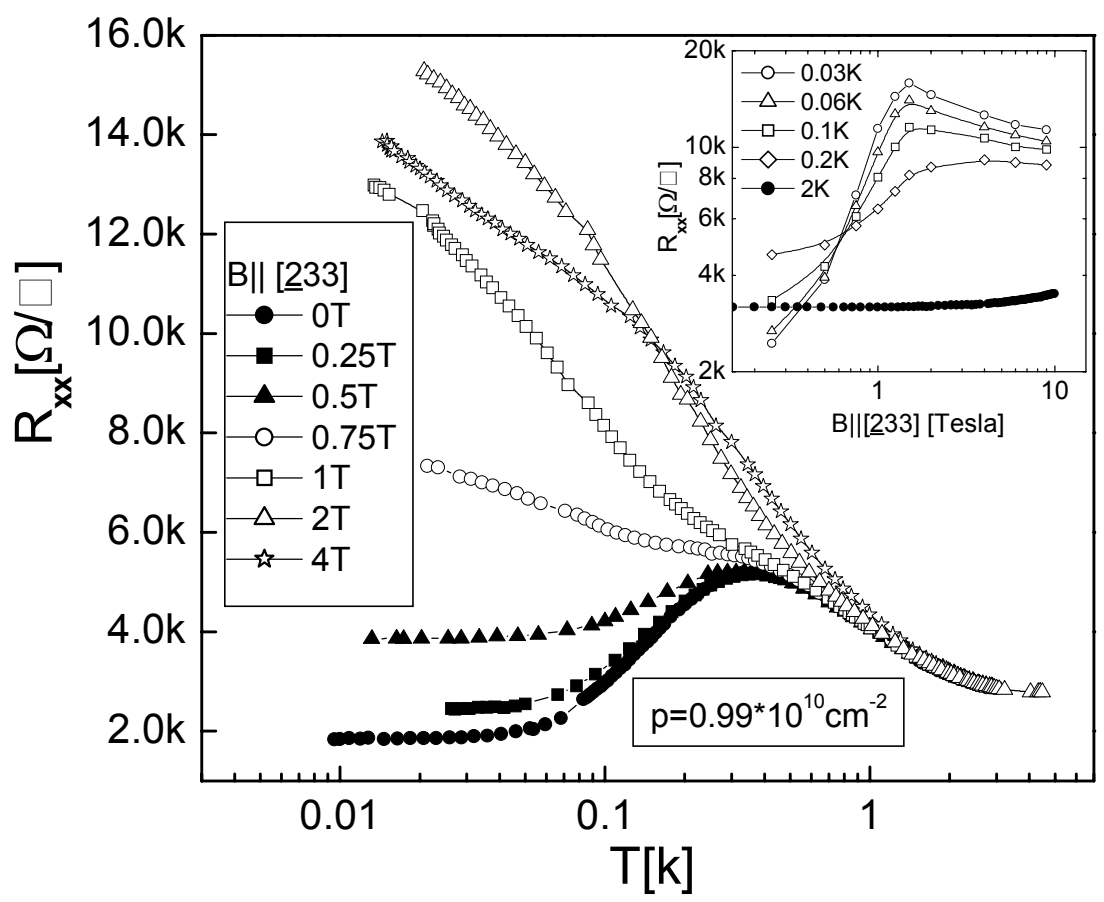

FIG. 2. $R_{xx}$ vs. $T$ at various $B_{||}$ for $p$=0.99×10$^{10}$ cm$^{-2}$. $B_{||}$ was along the [$\underline{2}$33] direction, the same direction as the current. It was determined that $B_{||}$ was tilted from the 2DHG plane about 0.1° by a Hall resistance measurement. The insert shows the isothermal magneto-resistance at different temperatures.

The 2D metallic state is known to be unstable against an in-plane parallel magnetic field [2]. For a high mobility GaAs heterostructure, a parallel field driven MIT with critical field $B_c$ was observed [8]. Our magneto transport data for density $p$=0.99×10$^{10}$ cm$^{-2}$ are presented in Fig. 2. The behavior for other densities is qualitatively similar.

There are several points we would like to address here. First, the parallel magnetic field $B_{||}$ has very small effect at temperatures higher than $T_0$, while at temperatures lower than $T_0$, $B_{||}$ suppresses the metallic behavior and the insulating behavior is 'restored' for $B_{||}>B_c$. This suggests that the metallic-like temperature dependent resistance in our system is a finite temperature effect. There is thus no indication of quantum phase transition, in contrast to high mobility Si-MOSFET's [3].

Our second point is that the field driven 'metal-insulator transition' is qualitatively consistent with a percolation transition where a metallic phase coexists with an insulating phase [6,13]. In Ref. 6, the 2D system is modeled as a mixture of a metallic liquid phase and an insulating phase. A parallel magnetic field vaporizes some of the liquid into the low density state, thus decreasing the overall conductivity. In Ref. 13, Meir modeled the 2D system as a percolation network of classical metallic and insulating quantum point contacts. Since the destruction of the metallic phase seems to be driven by the Zeeman energy, we believe that percolation in a coexisting liquid-gas model is more suitable for our sample. In particular, the $R_{xx}(T)$ curves at $B_{||}$=0.75 T and 1 T indicate the system is close to its percolation threshold at those fields.

Our third point is that at $T<T_0$, the isothermal magnetoresistance is negative at some field $B_{||} > B_c$. This nonmonotonic low temperature magnetoresistance is observed when the angle between the 2DHG plane and $B_{||}$ is very small. We take the negative magnetoresistance for $T<T_0$ and $B_{||}>B_c$ as evidence for the suppression of the weak localization effect by the small increasing perpendicular magnetic field due to the small misalignment between $B_{||}$ and the 2DHG plane.

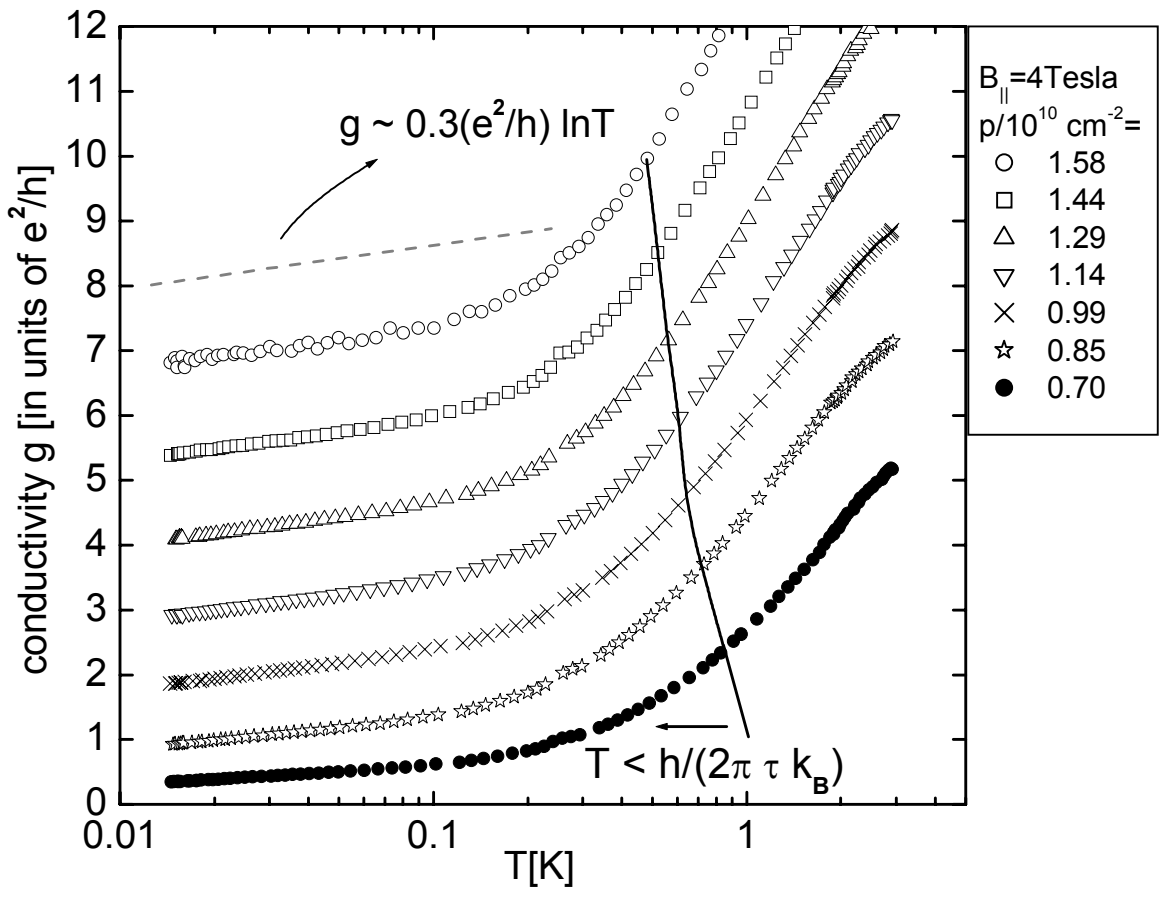

FIG. 3. Temperature dependent conductivity $g(T)$ of the 2DHG at various densities with a 4 T in-plane magnetic field along the [$\underline{2}$33] direction. The perpendicular field is roughly 70 G estimated from the Hall resistance. The dotted gray line represents $g(T) \sim 0.3(e^2/h)\ \ln T$, the function which approximately describes all the low temperature data for $g>e^2/h$. The solid black line connects the temperatures below which the transport become diffusive.

In Fig. 3, we present the temperature dependent conductivity $g(T)$ at 4T in-plane magnetic field along the [$\underline{2}$33] direction, under which the metallic resistance is well suppressed. It can be seen that for the upper six curves with $g(T)>e^2/h$, all the $g(T)$ traces can be approximately described by function $0.3(e^2/h)\ln T$ over almost a decade of temperature, 14mK-0.14K. On the other hand, the conductivity drops faster than $0.3(e^2/h)\ln T$ for the curve with lowest conductivity. This signals that different conduction mechanisms are dominating for $g > e^2/h$ and $g < e^2/h$. From the established weak localization theory, the low temperature conductivity correction from quantum interference for our symmetrically doped GaAs QW at small constant perpendicular field can be approximated as [14-16]

$$\Delta g(T) \approx (1/\pi)\ e^2/h\ \ln(T/T_\varphi), \qquad (1)$$

where $T_\varphi$ is the characteristic temperature below which the logarithmic temperature dependent conductivity from quantum interference starts. Eq. (1) is in quantitative agreement with the low $T$, $g>e^2/h$ data we show in Fig. 3. $T_\varphi$ depends on the ratio of the phase breaking time $\tau_\varphi$ and the momentum relaxation time $\tau$ of particles (holes). We do not have information on $\tau_\varphi$ of our system and therefore we do not know whether the systematic logarithmically decreasing conductivity for all the $g>e^2/h$ curves on Fig. 3 happens at the same temperature as $\tau_\varphi>\tau$ or not. As a comparison, we mark the temperature range $T< h/(2\pi\tau k_B)$ where transport becomes diffusive. It can be viewed as an upper limit of $T_\varphi$. In principle, since the weak localization suppression by a perpendicular field is well known, more insight can be gained by carefully tuning the angle between $B_{||}$ and the sample to introduce an adjustable small perpendicular field at constant $B_{||}>B_c$ to avoid the mixing of magnetoresistance induced by $B_{||}$ and the negative magnetoresistance due to weak localization. It will be a hard experiment and extreme care must be taken though, as several hundred gauss of perpendicular field may be enough to destroy the weak localization effect for a 2DHG in a high mobility GaAs quantum well.

The consistency between weak localization and low temperature transport of the 2DHG at high $B_{||}$ suggests that the insulating phase is dominated by the conductance of a usual Fermi liquid. Thus our data indicate the metallic state of our 2DHG system consists of a metallic liquid phase and a usual Fermi liquid. Local compressibility measurements by Ilani *et al.* reveal that a compressible background coexists with a phase that produces discrete screening events in the 2D metallic state [17]. This agrees with the picture here. The two-phase picture here may also naturally explain the continuous observation of a negative low field magnetoresistance in the 2D metallic state [18, 19]. More theoretical efforts are definitely needed.

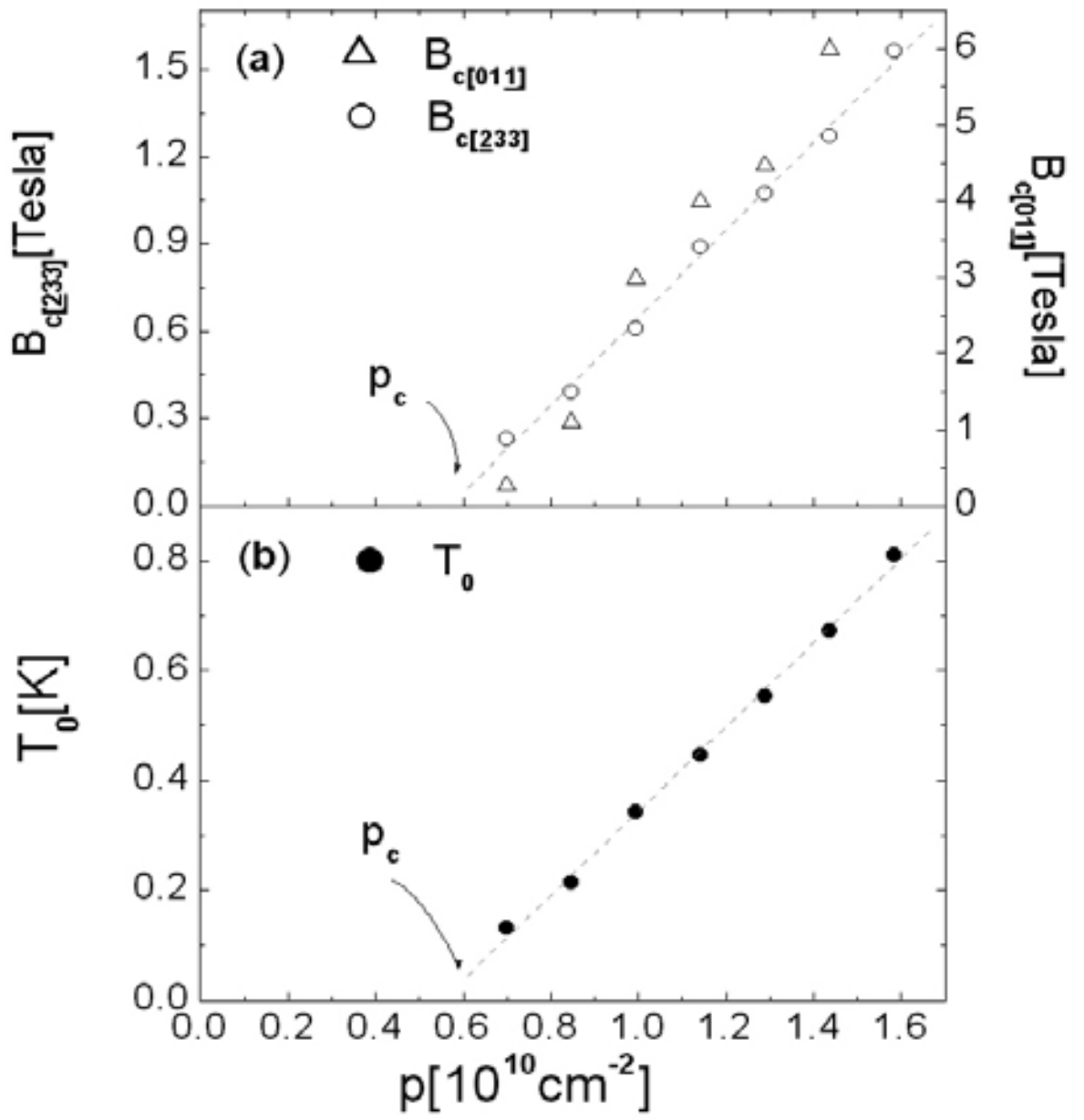

FIG. 4. (a) $B_c$ vs. $p$ for $B$ along the [01$\underline{1}$] and [$\underline{2}$33] directions. $B_c$ is determined from the crossing point of the low temperature isothermal magnetoresistance. Note that $B_{c[01\underline{1}]}$ should have larger error bars than $B_{c[\underline{2}33]}$ since for the [01$\underline{1}$] configuration, $B_{||}$ was tilted from the 2DHG plane by a relatively large angle, 2°. $B_{||}$ was only tilted from the 2DHG plane for 0.1° for the [$\underline{2}$33] configuration. (b) $T_0$ vs. $p$, where $T_0$ represents the temperature where the metallic resistance emerges as shown in Fig. 1. The dotted lines in (a) and (b) are guides to the eye.

The density dependences of the critical field $B_c$ for the [01$\underline{1}$] and the [$\underline{2}$33] directions are plotted in Fig. 4a. It is obvious that $B_c \propto (p\text{-}p_c)$ for both directions. Furthermore, the ratios of $B_c$ for the two directions are close to the ratio of expected anisotropic $g$ factor in GaAs QW's [20]. In Fig. 4b the $T_0$'s are plotted as function of 2DHG density. $T_0$ also extrapolates linearly to zero at the critical density. Assuming the $g$ factor is constant for all $p>p_c$, a natural interpretation for $B_c$, $T_0 \propto (p\text{-}p_c)$ is that the metallic state is instable as the Zeeman energy becomes greater than $E_a$, the energy scale associated with the metallic behavior. This is consistent with the two-phase separation theory [6,7] in which $E_a$ represents the cohesive energy of the liquid droplets. It is important to note that the product of the $g$ factor and electron effective mass for 2DEG in Si-MOSFET was found diverging at the critical density and the divergence is associated with the ferromagnetic instability in such system [4]. However, the data here can be readily interpreted with a constant $g$ factor and effective hole mass for $p>p_c$. This suggests that the metallic behavior for holes in GaAs may not be associated with ferromagnetic instability.

In summary, we have studied the effects of a parallel magnetic field on a high mobility dilute 2DHG in a GaAs quantum well over an extended density range. We demonstrate that the parallel magnetic field driven metal-insulator transition is like a percolation transition at finite temperature. At $B_{||}>B_c$, we observed a universal temperature dependent conductivity $g(T) \sim 0.3(e^2/h)\ln T$ at low $T$ for $g>e^2/h$, which is reminiscent of weak localization. The data suggest a metallic liquid phase coexists with a weakly localized 'normal' Fermi liquid gas phase in the 2D metallic state for our sample. The linear dependence of $B_c$ on density can be readily understood in terms of a density independent $g$ factor and effective mass $m^*$ for all the densities $p>p_c$, in contrast to the diverging $gm^*$ at $n_c$ for electrons in Si-MOSFET. This may indicate the metallic phenomenon originate from different mechanisms for holes in GaAs and electrons in Si-MOSFET.

---

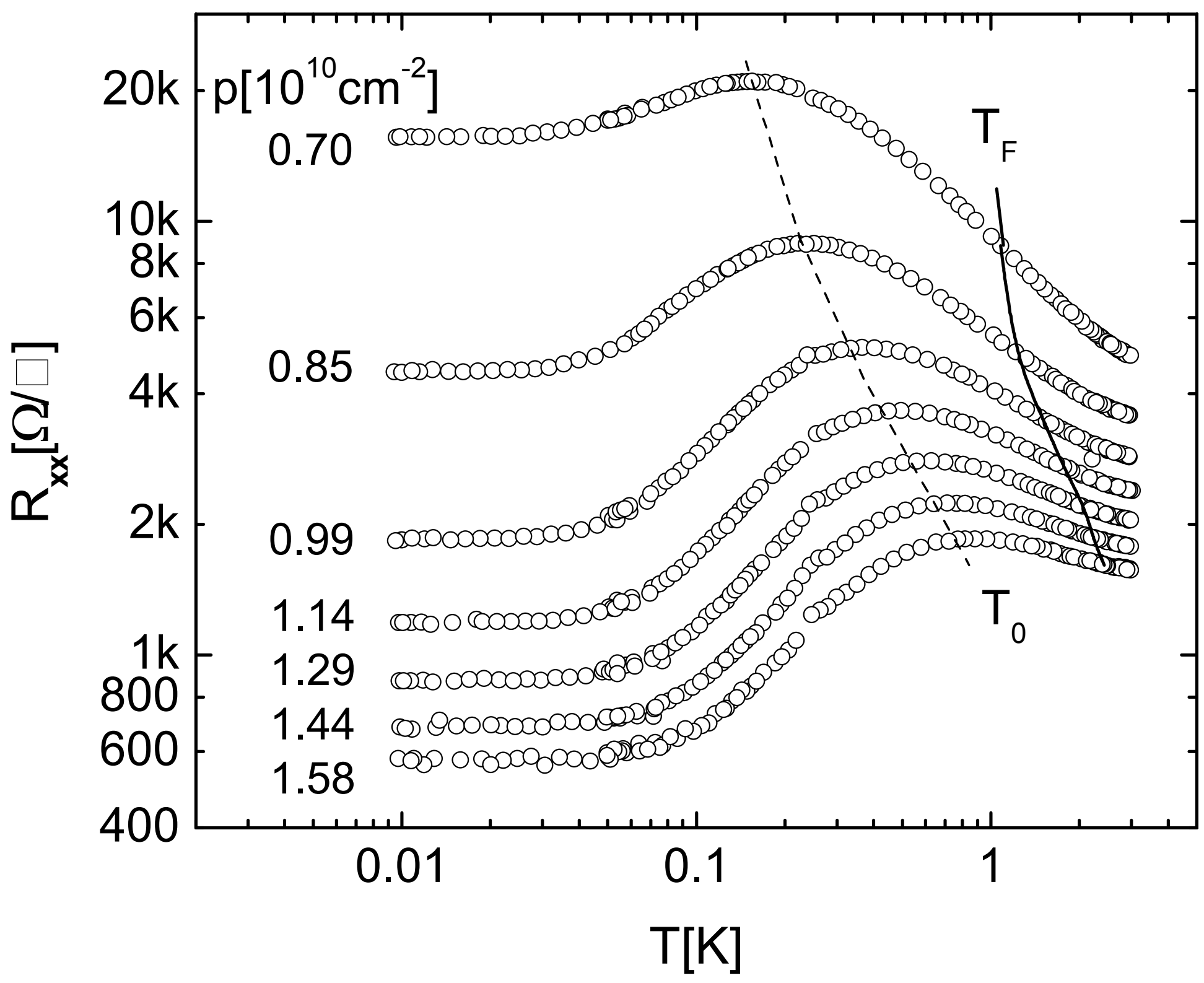

Fig. 1

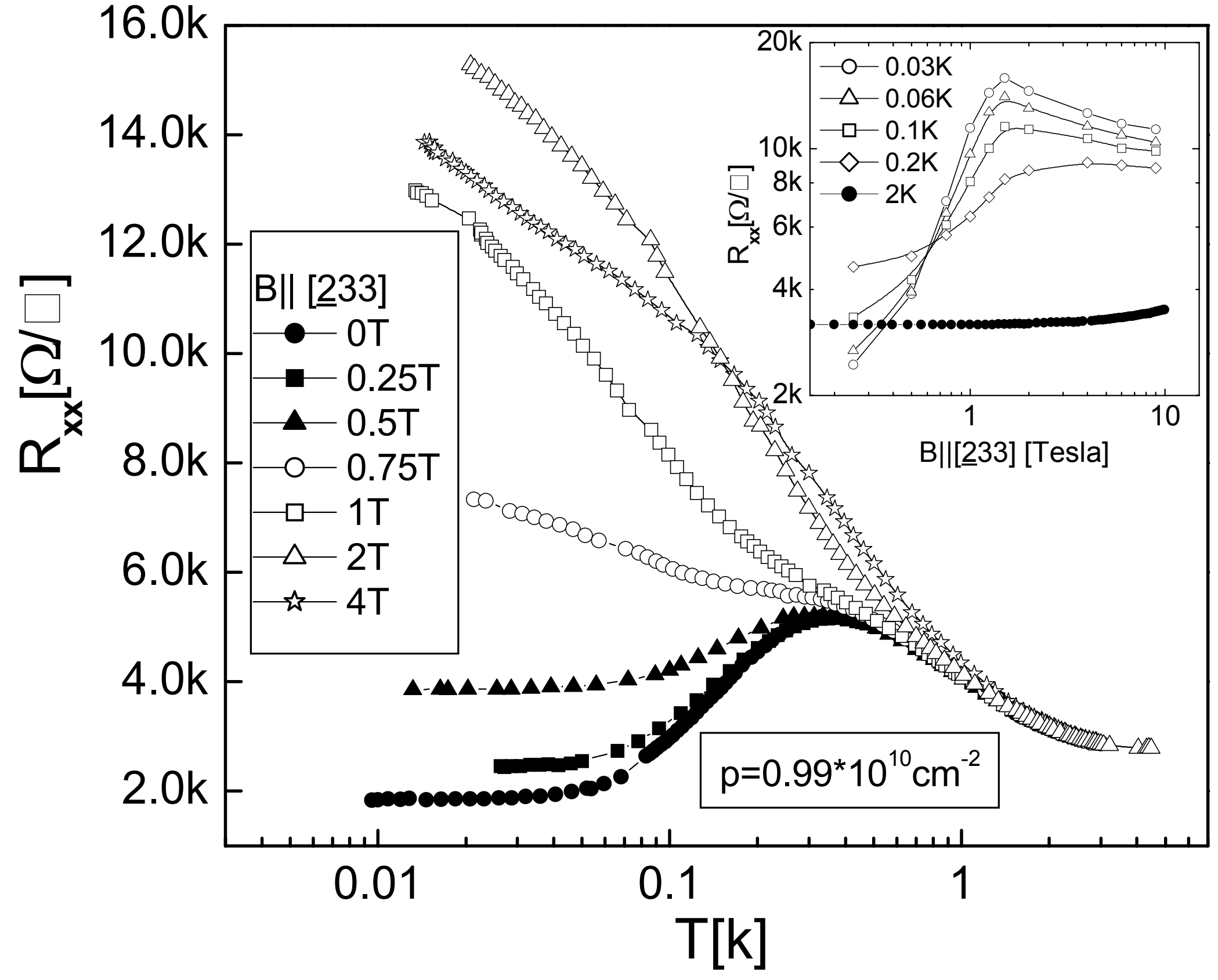

Fig. 2

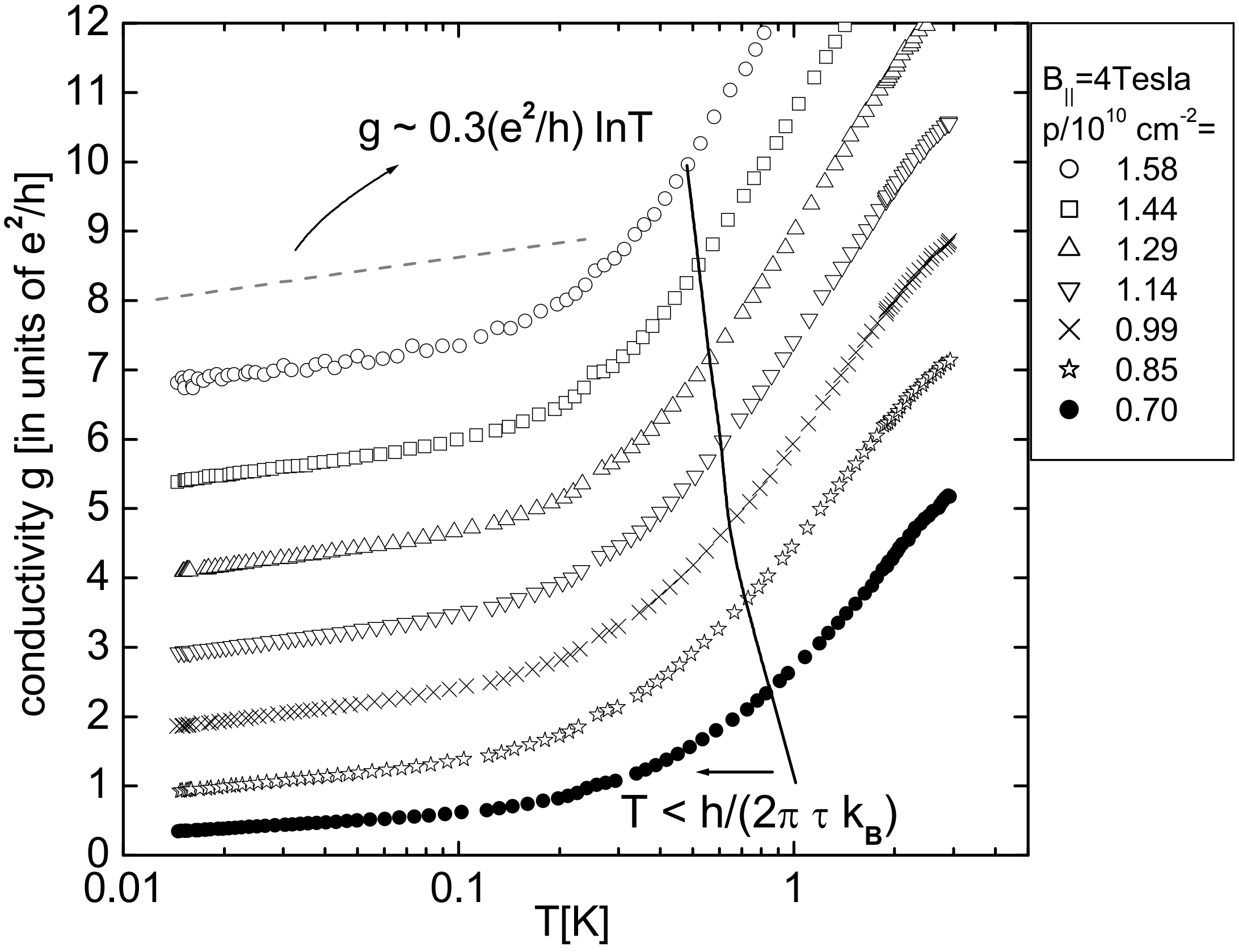

g ~ 0.3(e²/h) lnT
T < h/(2π τ k_B)
B_||=4Tesla
p/10¹⁰ cm⁻²=
1.58
1.44
1.29
1.14
0.99
0.85
0.70
conductivity g [in units of e²/h]
T[K]

Fig. 3

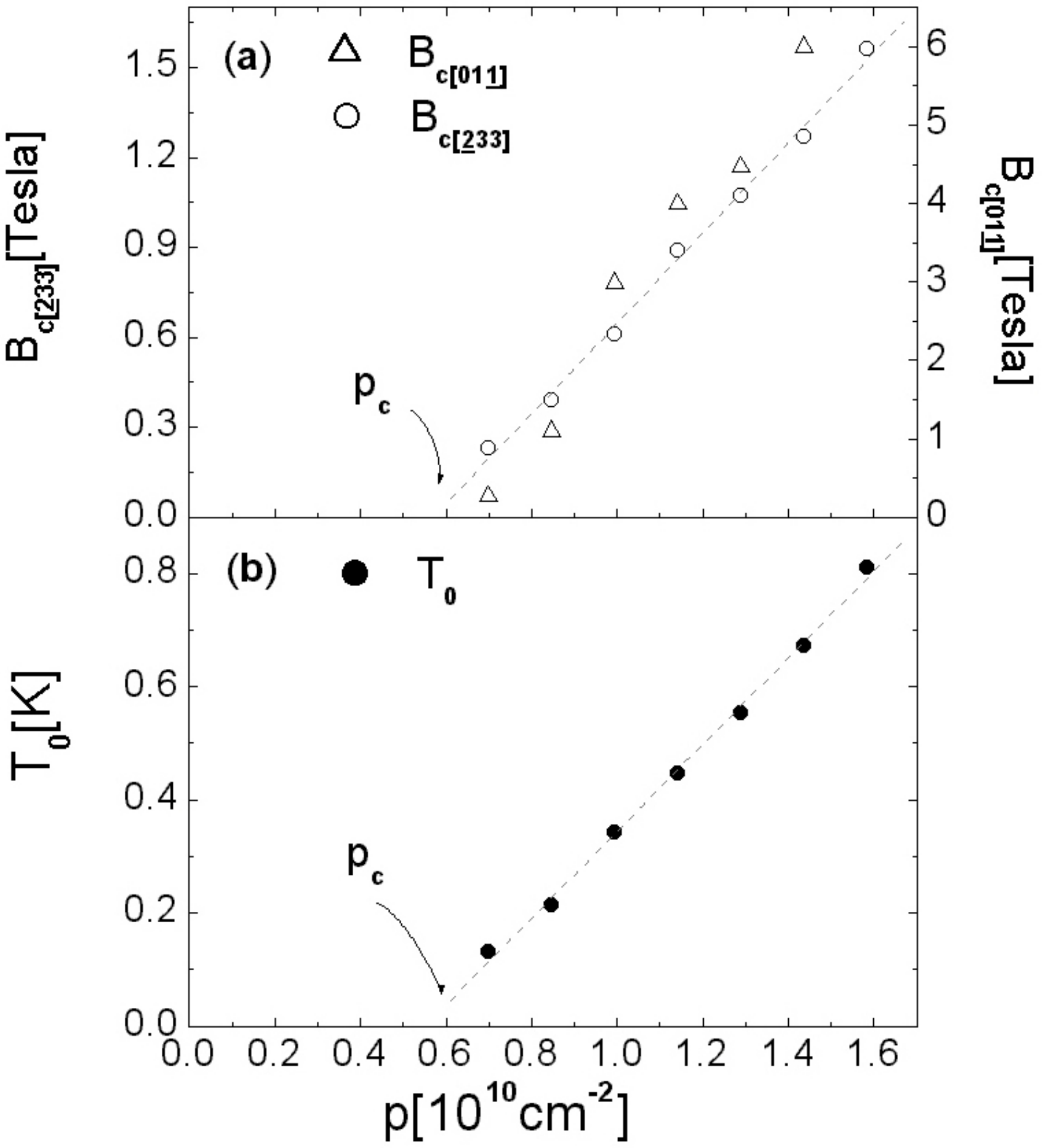

Fig. 4